\documentclass[aps,showpacs,onecolumn,preprint,amsmath,amssymb,superscriptaddress,intlimits,longbibliography,showkeys]{revtex4-1}

\usepackage{graphicx}
\usepackage{subfigure}
\usepackage{xcolor}
\usepackage{epstopdf}
\usepackage{amssymb}
\usepackage{amsmath}
\usepackage{amsfonts}
\usepackage{dcolumn}
\usepackage{bm}
\usepackage{soul}
\usepackage{array}
\usepackage{physics}

\newcommand{\vk}{{\bf k}}
\newcommand{\vQ}{{\bf Q}}
\newcommand{\vq}{{\bf q}}

\newcolumntype{C}[1]{>{\centering\let\newline\\\arraybackslash\hspace{0pt}}m{#1}}

\graphicspath{{figures/}}

\begin{document}

\title{Exciton thermalization dynamics in monolayer MoS$_2$: a first-principles Boltzmann equation study}

\author{Yang-hao Chan}
\email{yanghao@gate.sinica.edu.tw}
\affiliation{Institute of Atomic and Molecular Sciences, Academia Sinica, Taipei 10617, Taiwan}
\affiliation{Physic Division, National Center of Theoretical Physics, Taipei 10617, Taiwan}

\author{Jonah B. Haber}
\affiliation{Department of Physics, University of California, Berkeley, CA, 94720-7300, USA}
\affiliation{Materials Science Division, Lawrence Berkeley National Laboratory, Berkeley, CA, 94720, USA}

\author{Mit H. Naik}
\affiliation{Department of Physics, University of California, Berkeley, CA, 94720-7300, USA}
\affiliation{Materials Science Division, Lawrence Berkeley National Laboratory, Berkeley, CA, 94720, USA}

\author{Steven G. Louie}
\affiliation{Department of Physics, University of California, Berkeley, CA, 94720-7300, USA}
\affiliation{Materials Science Division, Lawrence Berkeley National Laboratory, Berkeley, CA, 94720, USA}

\author{Jeffrey B. Neaton}
\affiliation{Department of Physics, University of California, Berkeley, CA, 94720-7300, USA}
\affiliation{Materials Science Division, Lawrence Berkeley National Laboratory, Berkeley, CA, 94720, USA}

\author{Felipe H. da Jornada}
\email{jornada@stanford.edu}
\affiliation{Department of Materials Science and Engineering, Stanford University, Stanford, CA 94305, USA}

\author{Diana Y. Qiu}
\email{diana.qiu@yale.edu}
\affiliation{Department of Mechanical Engineering and Materials Science, Yale University, New Haven, CT 06520}

\begin{abstract}
Understanding exciton thermalization is critical for optimizing optoelectronic and photocatalytic processes in many materials. However, it is hard to access the dynamics of such processes experimentally, especially on systems such as monolayer transition metal dichalcogenides, where various low-energy excitations pathways can compete for exciton thermalization. Here, we study exciton dynamics due to exciton-phonon scattering in monolayer MoS$_2$ from a first-principles, interacting Green's function approach, to obtain the relaxation and thermalization of low-energy excitons following different initial excitations at different temperatures.
We find that the thermalization occurs on a picosecond timescale at 300 K but can increase by an order of magnitude at 100 K. 
The long total thermalization time, owing to the nature of its excitonic band structure, is dominated by slow spin-flip scattering processes in monolayer MoS$_2$.
In contrast, thermalization of excitons in individual {spin-aligned} and spin-anti-aligned channels can be achieved within a few hundred fs when exciting higher-energy excitons. 
We further simulate the intensity spectrum of time-resolved angle-resolved photoemission spectroscopy (TR-ARPES) experiments and anticipate that such calculations may serve as a map to correlate spectroscopic signatures with microscopic exciton dynamics.
\end{abstract}

\date{\today}
\maketitle

\section{Introduction}

Upon exciting a material with light, photogenerated excitons -- correlated electron-hole pairs  -- can undergo a series of scattering processes before the constituent electrons and holes recombine and emit light.  
A thorough understanding of these processes is not only critical for the fundamental understanding of the dynamics of photoexcited carriers {in quasi two dimensional (2D) systems with strongly bound excitons}, but also for applications in optoelectronics and photocatalysis~\cite{Mueller2018}.

The study of exciton dynamics with optical pump-probe experiments has proven to be critical for the interpretation of excited-state properties of 2D materials and their ability to convert photons to other sources of energy\cite{Wang2013,Mai2014,Lagarde2014,Poellmann2015,Wang2018}.
In particular, optical characterization studies on heterostructures of quasi-2D materials reveal ultrafast exciton transfer time~\cite{Hong2014,Chen2016,Jin2018}, and 
analyses of time-resolved photoluminescence spectra are indispensable tools to extract exciton recombination times along with other exciton scattering and dephasing processes~\cite {Lagarde2014,Zhang2015,Robert2016}. This is particularly relevant as the community considers excitons as a potential quantum state that can be used to store quantum information and generate single photons for quantum communication applications.
However, it is not straightforward to directly interpret experimental quantities (such as exciton linewidth) as exciton scattering times, especially when such interpretation often resorts to simulations of kinetic equations with a number of adjustable parameters \cite{Mak2012,Kioseoglou2012,Yoon2022}.
The lack of a practical, microscopic framework to access the dynamics of photoexcited carriers and the dynamics of both bright and dark excitons further complicates the interpretation of experimental observables. Such a knowledge is particularly critical with the availability of new experimental capabilities, such as those based on momentum microscopes\cite{Madeo2020,Man2021,Kunin2023}, to directly measure the dynamics of excited states in energy and momentum spaces.

Exciton dynamics in low dimensional transition metal dichalcogenides (TMD) are of particular interests owing to the rich physics from their electronic structure with intricate valley-spin couplings and their experimental accessibility due to the large binding energy of excitons. First-principle simulations of exciton dynamics in monolayer TMD based on a combination of time-dependent density functional theory (TD-DFT) and non-adiabatic molecular dynamics (NAMD) approaches was conducted~\cite{Liu2020,Jian2021} with excited-state potential energy surfaces from either range-separated hybrid functionals or many-body perturbation theory by solving the Bethe-Salpeter equation (BSE).
Valley depolarization dynamics and photo-excited (single-particle) carriers relaxation path were revealed by \textit{ab initio} studies based on nonequilibrium Green function approach~\cite{Molina-Sanchez2017}.
In exciton based approaches, exciton relaxation dynamics was studied by solving a coupled equation of motion of exciton polarization and exciton-exciton correlations for a two-dimensional TMD model~\cite{Brem2018}.
Alternatively, dynamics of exciton populations in an incoherent limit can be described by an exciton Boltzmann equation~\cite{Snoke1991,Snoke2011,Snoke2014}.
A recent first-principles calculation of exciton Boltzmann equation in monolayer WSe$_2$ predicts a valley depolarization time of hundred femtoseconds and simulates transient absorption and TR-ARPES spectrum~\cite{Chen2022}.

In this paper, we study exciton dynamics in monolayer MoS$_2$ by numerically solving an exciton Boltzmann equation from first-principles.
We show that exciton thermalization time is strongly dependent on temperature and initial excitation conditions.
A full thermalization for resonant excitons and close-to-band-edge excitations takes about 1 ps and 20 ps at 300 K and 100 K, respectively. The long thermalization time stems from slow spin-flip scattering mediated by chiral phonons~\cite{Zhang2022}.
Owing to the much faster spin-conserving scattering rate, thermalization in the spin-aligned or spin-anti-aligned channel can be reached on a shorter time scale. This can be clearly seen when excitation energy is at resonance with the A excitons. With close-to-band-edge excitations the separation in thermalization time is less clear.
To complement our detailed calculations of energy and momentum resolved exciton population dynamics, we also simulate TR-ARPES intensity spectra, revealing important spectral features at different stages of the thermalization process such as fast valley transfer and flattening of spectrum dispersion over time.

The rest of the paper is organized as follows. In Sec.~\ref{sec:method}, we introduce the Boltzmann equation and our implementation of the numerical solver. 
We benchmark our implementation by comparing extracted relaxation times against relaxation times computed from many-body perturbation theory (MBPT). Details of the first-principles calculations are also given.
In Sec.~\ref{sec:popdynamics}, we analyze the temperature and the initial condition dependence of the low energy exciton population dynamics and study distinct scattering paths associated with the two different, but experimentally relevant, initial conditions. 
In Sec.~\ref{sec:trARPES}, we simulate the intensity spectrum of TR-APRES experiments, which demonstrates key features and time scales of exciton transfer processes.
Sec.~\ref{sec:thermalization} is devoted to the investigation of the thermalization process and analysis of how a Boltzmann distribution is reached.
We find a clear separation in time scales of intra-species from inter-species exciton thermalizations, which is ascribed to the internal spin structure of excitons in monolayer MoS$_2$.
A conclusion is given in Sec.~\ref{sec:conclusion}.

\section{Method}
\label{sec:method}

\begin{figure}[t]
     \centering
     \includegraphics[width=.48\textwidth]{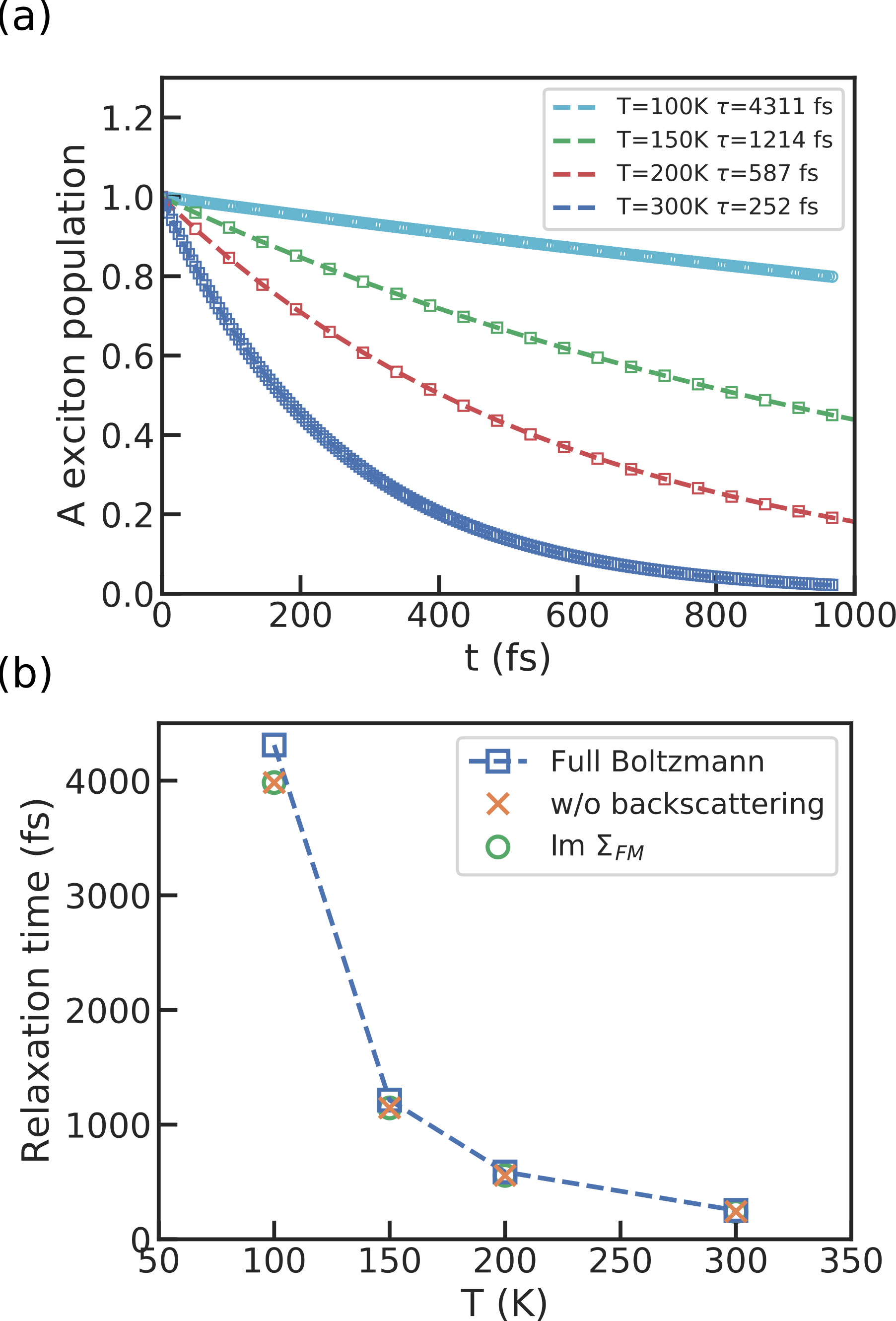}
\caption{(a) Relaxation time of exciton A extracted from Boltzmann dynamics simulations using Eq. (1) at different temperature. (b) Extracted relaxation time from simulated dynamics with (blue squares) and without back-scattering terms (orange crosses) compared with those computed from the imaginary part of the Fan-Migdal self-energy (green circles).}
\label{fig:Apop}
\end{figure}

The Boltzmann equation for the exciton population, $f_{n\vQ}$, of an exciton state labeled by $n$ and center-of-mass momentum (COM) $\vQ$ can be derived from the cluster expansion technique~\cite{Kira2006} or many-body perturbation theory~\cite{Snoke2011}. 
Assuming a homogeneous real-space distribution without external forces, dynamics of exciton populations are governed by,
\begin{widetext}
\begin{align}
\frac{\partial}{\partial t}f_{n\vQ}(t) =
 \frac{2\pi}{\mathcal{N}_\mathbf{q}\hbar}\sum_{\mathbf{q},l,\nu} \left|g_{ln\vQ}^{\mathbf{q}\nu}\right|^{2}\left\{\right.&-f_{n\vQ}\left(1+f_{l\vQ+\mathbf{q}}\right) \left[n_{\mathbf{q}\nu}D^{+}_{nl\vQ\mathbf{q}} 
 +\left(1+n_{\mathbf{q}\nu}\right)D^{-}_{nl\vQ\mathbf{q}}\right]\nonumber\\
 & \left.+f_{l\vQ+\mathbf{q}}\left(1+f_{n\vQ}\right)\left[(n_{\mathbf{q}\nu}+1)D^{+}_{nl\vQ\mathbf{q}}+n_{\mathbf{q}\nu}D^{-}_{nl\vQ\mathbf{q}}\right]\right\},
 \label{eq:boltz}
\end{align}
\end{widetext}
where the right-hand side is the collision term which describes how band and momentum resolved exciton populations evolve in real-time due to microscopic exciton-phonon scattering events. In this expression, $n_{\mathbf{q}\nu}$ is the phonon occupation of a mode $\nu$ and crystal momentum $\mathbf{q}$, $\mathcal{N}_\mathbf{q}$ is the total number of $\mathbf{q}$ points and $g^{\mathbf{q}\nu}_{ln\vQ}$ denotes the exciton-phonon coupling matrix element, which describes the coupling of an exciton of state $(n,\vQ)$ to another state $(l,\mathbf{Q+q})$ via a $(\nu,\mathbf{q})$ phonon~\cite{Chen2020,Antonius2022}. 
The energy conservation of scatterings is guaranteed through $D^{\pm}_{nl\vQ\mathbf{q}}=\delta(-\epsilon_{l\vQ+\mathbf{q}}+\epsilon_{n\vQ}\pm\hbar\omega_{\mathbf{q}\nu})$, where $\epsilon_{n\vQ}$ is the exciton excitation energy and $\omega_{\mathbf{q}\nu}$ is the phonon energy. 
The physical content of each term in the collision term is clear: the first two terms describe excitons being scattered from the state $(n,\vQ)$ to the other state $(l,\mathbf{Q+q})$ mediated through the absorption and emission of a phonon, respectively. The $\left(1+n_{\mathbf{q}\nu}\right)$ factor emphasizes the bosonic enhancement effect of the phonons. The remaining two terms in the second line describe an increase in the occupation of state $(n,\vQ)$ population by scatterings initiated from the state $(l,\mathbf{Q+q})$.  
In deriving Eq.~\ref{eq:boltz}, we make a Markov approximation and further assume that excitons are of bosonic nature, which is valid in the dilute limit. The populations of phonons are assumed to follow a Bose distribution at a fixed temperature, which treats phonons as a bath, ignoring their dynamics. Exciton coherences and higher order correlations are also ignored.

We implemented a numerical solver for Eq.~\ref{eq:boltz}. All the required ingredients, such as phonon modes, exciton energies, and exciton-phonon coupling matrix elements, are computed from first-principles~\cite{Chan2023}. The time-propagation is integrated with a fourth order Runge-Kutta algorithm with a step size of 0.5 fs.
We include 20 exciton bands in the calculations. 
Numerically, delta functions are implemented as Gaussian functions with widths determined by the adaptive smearing method~\cite{Yates2007,Li2014,Cepellotti_2022}. 
The only free parameter left is an overall prefactor in the adaptive smearing method, which we set to be 0.5 on which the qualitative features are relatively insensitive. 

Computational details of first-principles calculations are given as follows. Electronic structure and density functional perturbation theory~\cite{Baroni2001} calculations are performed with the Quantum Espresso package~\cite{qtespresso}.
Electron-phonon coupling matrix elements are computed with the EPW package~\cite{PONCE2016}.
We compute GW quasi-particle energy and solve for excitons with the BerkeleyGW package~\cite{Hybertsen1986,Rohlfing2000,Deslippe2012}. 
Finite COM excitons~\cite{Qiu2015} and phonons are both computed on a $48\times48$ $\vQ$-grid and $\vq$-grid. Exciton-phonon coupling matrix elements are computed on the same grids then linearly interpolated to a $96\times96$ $\vQ$- and $\vq$-grid while both phonon and exciton energies are interpolated with a cubic polynomial when solving Eq.~\ref{eq:boltz} for better convergence.
All other parameters can be found in our previous work~\cite{Chan2023}.

One can show that Eq.~\ref{eq:boltz} conserves the total exciton populations, which we also check numerically. To further benchmark our implementation, focusing on the lowest energy bright exciton (commonly denoted as the A exciton), we compare relaxation times extracted from our real-time simulation with relaxation times computed from the imaginary part of Fan-Migdal (FM) like exciton-phonon self-energy~\cite{Antonius2022,Chen2020},
\begin{equation}
\frac{1}{\tau_{n\vQ}} = \frac{2\pi}{\hbar\mathcal{N}_\mathbf{q}}\sum_{m,\vq,\nu,\pm}|g_{nm\vQ}^{\vq\nu}|^2(n_{\nu\vq} +\frac{1}{2}\pm\frac{1}{2})\delta(\epsilon_{n\vQ}-\epsilon_{m\vQ+\vq}\mp\hbar\omega_{\nu\vq}).
\label{Eq:tau}
\end{equation}
To extract the relaxation time of the A exciton, we fix the temperature and start the simulation with a single A exciton and subsequently monitor the population changes as a function of time. The proceedure is repeated at different temperatures as shown in Fig.~\ref{fig:Apop} (a). The trajectory can be well-fitted with an exponential function $e^{-t/\tau_A(T)}$ to extract $\tau_A(T)$. In Fig.~\ref{fig:Apop} (b) we plotted the extracted relaxation times together with the perturbative FM results.
The results from solving Boltzmann equation (blue squares) slightly deviated from FM results at low temperature. 
The deviation can be understood by realizing that Eq.~\ref{Eq:tau} can also be derived by setting $f_{l\mathbf{Q+q}}=0$ in the right-hand side of Eq.~\ref{eq:boltz}. Therefore, the scatterings from all other states back to A exciton are ignored in FM. To verify this, we also simulate the dynamics with all the back-scattering terms set to zero. The extracted relaxation time (orange crosses) in Fig.~\ref{fig:Apop} (b) shows perfect agreement with the FM results, which validates our implementation.

\section{Exciton dynamics}
\label{sec:popdynamics}

\begin{figure}[t]
     \centering
     \includegraphics[width=.48\textwidth]{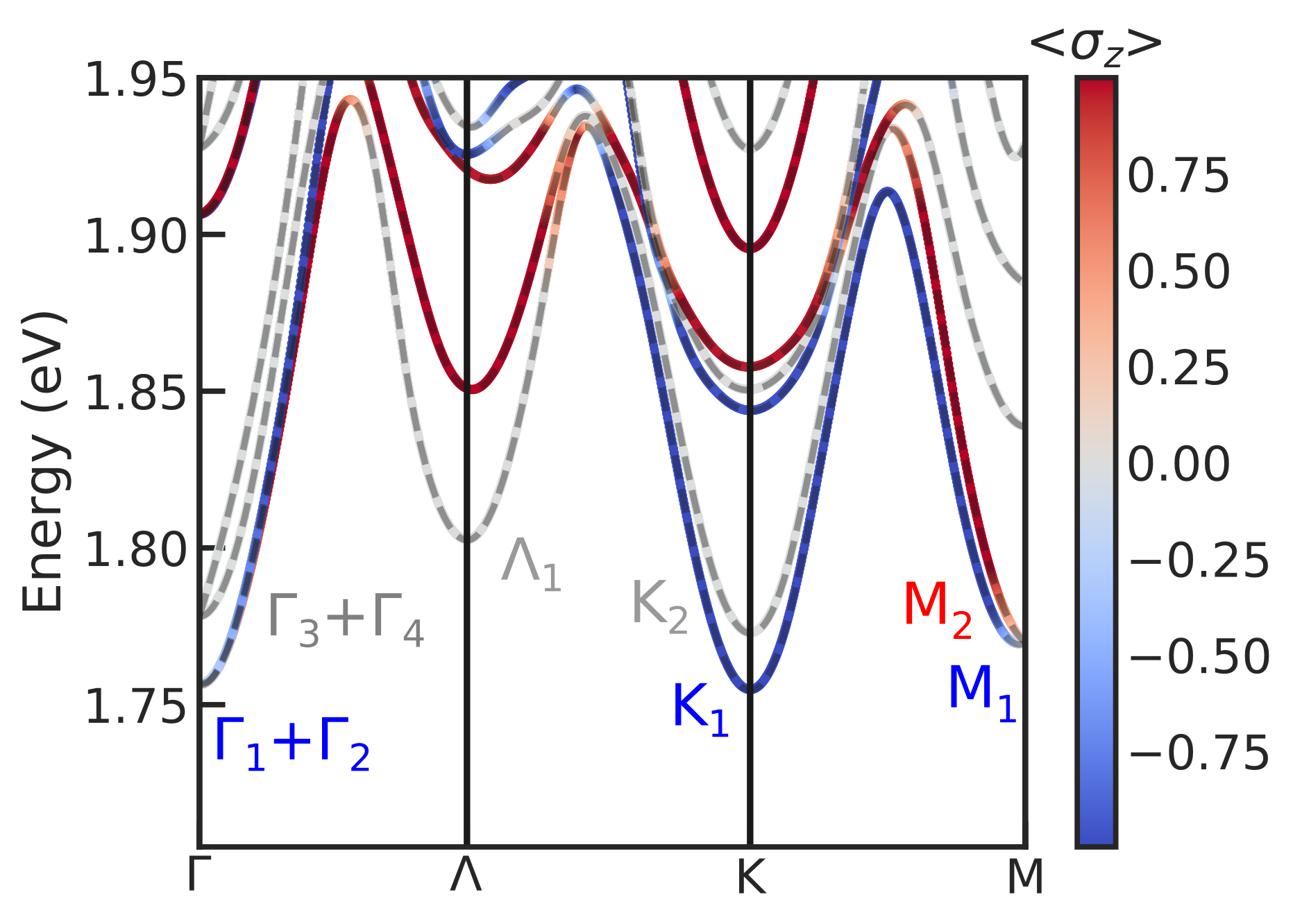}
\caption{Exciton dispersion (exciton band structure) in monolayer MoS$_2$. The color scale shows the spin expectation values of each exciton. Spin-anti-aligned exciton states are those with finite spin expectation values. We label excitons in low energy valleys by their COM and their order in energy at that valley.}
\label{fig:exciton_dispersion}
\end{figure}

We study exciton dynamics by tracking exciton populations at well-defined low energy valleys in the exciton band structure as excitons can spend a significant amount of time there after excitation.
One important feature of these low energy valleys in a monolayer MoS$_2$ is that the total electron-hole spin is almost a good quantum number for most COMs except for along the $\Gamma$-$M$ high symmetry line.
In Fig.~\ref{fig:exciton_dispersion}, we show the exciton dispersion for a monolayer MoS$_2$ overlaid with their spin expectation values, which is defined as
\begin{equation}
    \langle n\vQ\left|\sigma_z\right| n\vQ\rangle \equiv \sum_{cc'v\vk} A^{n,*}_{c\vk+\vQ,v\vk} \langle c\vk+\vQ\left|\sigma_{z}\right| c'\vk+\vQ\rangle A^{n}_{c'\vk+\vQ,v\vk} -\sum_{cvv'\vk} A^{n,*}_{c\vk+\vQ,v'\vk} \langle v\vk\left|\sigma_{z}\right| v'\vk\rangle A^{n}_{c\vk+\vQ,v\vk},
\end{equation}
where $A^{n}_{c\vk+\vQ,v\vk}$ is the exciton envelope function for an exciton with COM $\vQ$ and quantum number $n$ and $c$ and $v$ label electron conduction and valence bands, respectively.
We label excitons by their COM and their order in energy counted from the lowest energy state at that COM (e.g. the second lowest energy exciton with COM $K$ is labeled $K_2$). 
Exciton spin expectation values clearly indicate the nature of the spin alignment of the electron and hole, which we will denote as spin aligned (anti-aligned) for the electron and hole with anti-aligned (aligned) spin orientation, of the individual excitons.
For example, the optically bright A exciton, which is doubly-degenerate due to the valley degeneracy and labelled  as $\Gamma_3$ or $\Gamma_4$, is the second lowest energy state at the $\Gamma$ point and spin-aligned, while the lowest energy states are spin-anti-aligned dark states. 
Our previous work has shown that the internal spin structure can lead to selection rules for the exciton-phonon couplings~\cite{Chan2023} and plays an important role in exciton dynamics.
Since dynamics in separate channels are expected to be only weakly coupled, the spin-flipping process should be the bottleneck for reaching a fully thermalized distribution among excitons of different spin species.

\begin{figure*}[t]
     \centering
     \includegraphics[width=\textwidth]{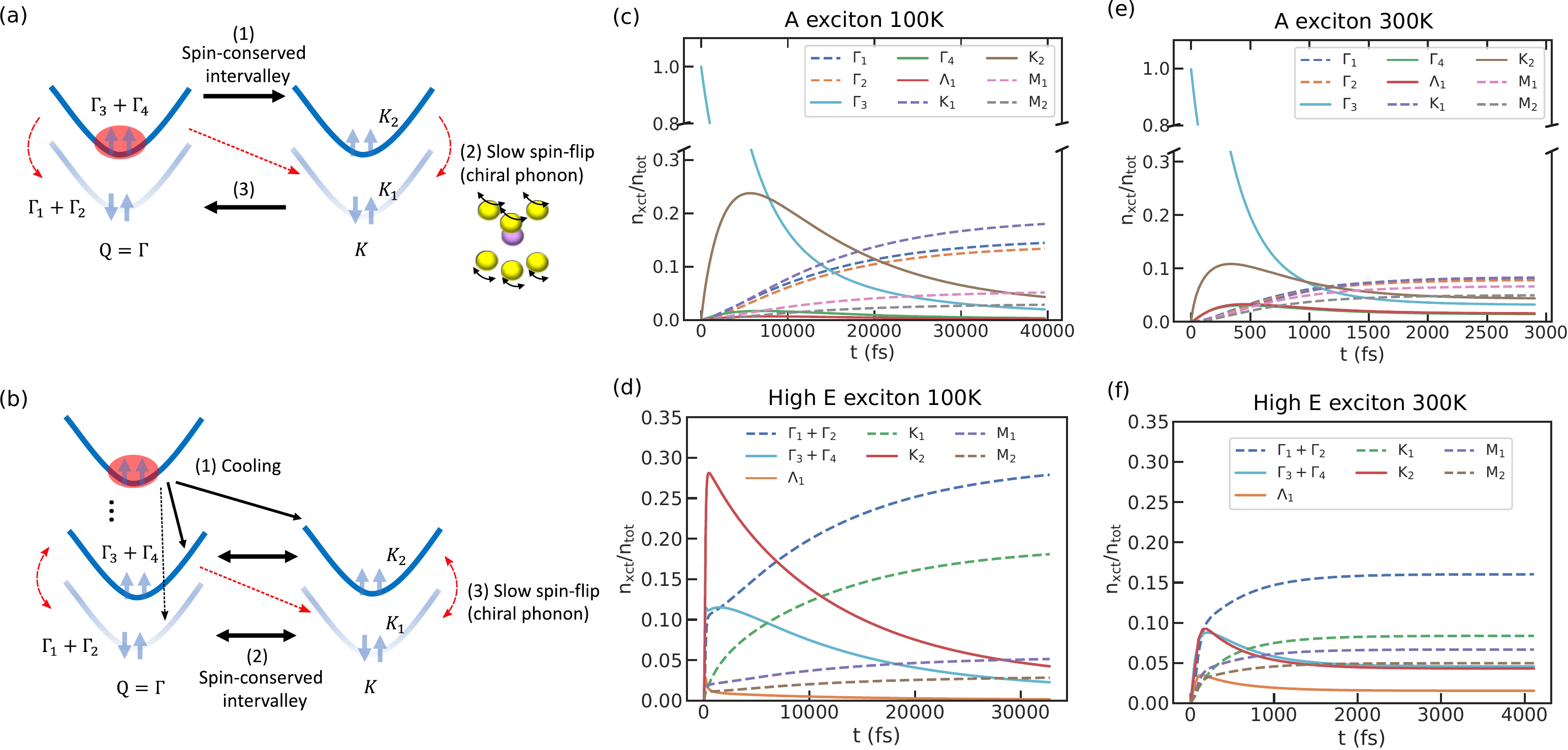}
\caption{Evolution of exciton populations at low energy valleys defined in Fig.~\ref{fig:exciton_dispersion} for four different initial setups. (a) and (b) are schematic plots of thermalization pathways at two different temperatures corresponding to the case (c), (e) and (d), (f), respectively. The two arrows indicate the relative orientation of the spin of the excited electron in the conduction band to that of the missing electron in the valence band. In (a) and (b) solid (dashed) lines indicate spin-conserved (spin-flip) scatterings. In (c) and (e), we start with a single exciton in one of the A exciton states at $\vQ=0$; in (d) and (f) a Gaussian distribution of high energy excitons centered at 2.05 eV is initialized. The temperature is 100 K for simulations (c) and (d) and 300 K for simulations (e) and (f), respectively. Dashed (solid) lines in panel (c), (d), (e), and (f) are used for populations of spin-anti-aligned (spin-aligned) excitons.}
\label{fig:popt}
\end{figure*}

The goal of this work is to understand exciton relaxation pathways and the dependence of relaxation dynamics on temperature and excitation conditions. For this purpose, we performed simulations with four different setups, which are
\begin{enumerate}
    \item a single A exciton initial state at 100 K,
    \item a single A exciton initial state at 300 K,
    \item an initial distribution of excitons centered at 2.05 eV with a width of 0.05 eV at 100 K,
    \item an initial distribution of excitons centered at 2.05 eV with a width of 0.05 eV at 300 K.
\end{enumerate}
Setup 1 and 2 simulate resonant excitations of the A exciton while setup 3 and 4 are relevant for a close-to-band-edge excitation in the quasiparticle band structure.
The resulting time-evolution of exciton populations in the low energy valleys for these four setups 1, 2, 3, and 4 is shown in Fig.~\ref{fig:popt} (c), (e), (d), and (f), respectively.
Unless stated otherwise, exciton populations reported below are summed over a patch of a 0.15 $\AA^{-1}$ radius centered at the labeled high symmetry COMs.

We first note the strong temperature dependence of the exciton dynamics.
At 300 K both simulations in Fig.~\ref{fig:popt} (e) and (f) reach steady states within 3 ps. In contrast, dynamics in (c) and (d) take more than 30 ps at 100 K to reach their steady states. 
Such an order of magnitude difference is consistent with the temperature dependence of A exciton relaxation time shown in Fig.~\ref{fig:Apop}, where we see that by decreasing the temperature from 300 K to 100 K, the relaxation time increases by more than 10 times.

Despite different time scales required to reach steady states at different temperatures, exciton population dynamics under the same initial excitation energy show qualitatively similar behaviors. 
For cases where we start with a $\Gamma_3$ exciton, we observe a clear rise and a subsequent slow decrease of the $K_2$ population in Fig.~\ref{fig:popt} (c) and (e).
Similar behaviors are also observed in $\Gamma_4$ and $\Lambda_1$ excitons. However, the increase in populations of $\Lambda_1$ excitons is smaller at 100 K due to its higher excitation energy.  
Populations of spin-anti-aligned excitons (represented by dashed lines), including the lowest two energy states $\Gamma_1+\Gamma_2$ and $K_1$, increase monotonically and compensating the decrease in populations of the $K_2$ and $\Gamma_3$ excitons.

We can understand these dynamics from the $\vQ$-space exciton energy landscape, the relatively slow relaxation time of $\Gamma_3$, $\Gamma_4$, and $K_2$ excitons, and the long spin-flip scattering time.
Shematic plots of primary relaxation pathways for the two different initial setups which populate the A exciton state at two different temperatures are given in Fig.~\ref{fig:popt} (a).
With the A exciton occupied, the $K_2$ excitons are quickly populated via spin-conserved intervalley scattering.
Relaxation to spin-anti-aligned excitons can only be achieved via spin-flip scatterings, which occurs at significantly slower rates~\cite{Zhang2022}.
For monolayer MoS$_2$, we find that there are two spin-flip channels within the low energy valleys.
One is the $K_2$ to $K_1$ scattering and the other is the scattering between $\Gamma_3$, $\Gamma_4$ and $\Gamma_1$, $\Gamma_2$, both of which are mediated by chiral phonons with a momentum $\vq\simeq0$, similar to that in WS$_2$~\cite{Zhang2022}. 
Our analysis shows that the scattering rate of the $K_2$ to $K_1$ channel is about one order of magnitude higher than that of the spin flip channel at the $\Gamma$ point, which dominates the spin-flipping process hence the total thermalization time.
Once spin-anti-aligned excitons are populated, relatively faster spin-conserved intervalley scattering between $K_1$ and $\Gamma_1$, $\Gamma_2$ can further thermalize the excitons.

When starting from high energy spin-aligned excitons, as in setups 3 and 4 described above, we see a sharp rise in the $K_2$ population in the first hundred fs. Populations of $\Gamma_3+\Gamma_4$ and spin-anti-aligned excitons $\Gamma_1$ and $\Gamma_2$ also increase quickly at the same time as shown  in Fig.~\ref{fig:popt} (d) and (f).
After the first hundred fs, populations of spin-aligned $K_2$, and $\Gamma_3+\Gamma_4$ excitons start a slow relaxation over 2 ps and 30 ps at 300 K and 100 K, respectively.
Such population trapping-like behavior is particularly clear for $K_2$ excitons in Fig.~\ref{fig:popt} (d). 
During the same time window, populations of spin-anti-aligned excitons including $K_1$, $\Gamma_1$, and $\Gamma_2$ excitons increase monotonically.
The fast cooling in the first hundred fs can be understood from the large scattering rates of high energy excitons.
Moreover, since $\Gamma_1$ and $\Gamma_2$ are spin-anti-aligned excitons, their increases in populations suggest that high energy excitons have significantly higher spin-flip scattering rates.
We find that the fast increase of populations of $\Gamma_1$ and $\Gamma_2$ excitons is due to efficient scatterings from excitons with COMs close to $M$, which are formed by electron at $\Lambda$ valley between $\Gamma$ and $K'$ valley and hole at $K$ valley. 
In contrast to the case with an initial $\Gamma_3$ exciton, populations of spin-anti-aligned excitons, $\Gamma_1$ and $\Gamma_2$ was comparable to those of spin-aligned excitons $\Gamma_3$ and $\Gamma_4$  when the slow relaxation began.
In the second stage, spin-conserved intervalley scattering equlibriates excitons in each channel.
In the mean time slow spin-flip scatterings drive the system to a full thermal distribution.
We will investigate how the thermalization is reached in Sec.~\ref{sec:thermalization} and show that exciton distributions over their energy at long time indeed follow the Boltzmann distribution.

\section{Simulation of TR-ARPES}
\label{sec:trARPES}

Although valley-resolved population dynamics shown in Fig.~\ref{fig:popt} may be deduced from optical measurements, the complicated exciton energy landscape and the intertwined interactions with defects and phonons pose great challenges to the analysis. On the other hand, TR-ARPES could provide direct access to exciton dynamics and has become a powerful technique to study exciton dynamics. In principle, TR-ARPES has resolution in both momentum and energy space and can detect both bright and dark excitons. In recent experiments, the first few ps exciton dynamics was reported for a monolayer WS$_2$~\cite{Madeo2020}. 

\begin{figure*}[t]
     \centering
     \includegraphics[width=\textwidth]{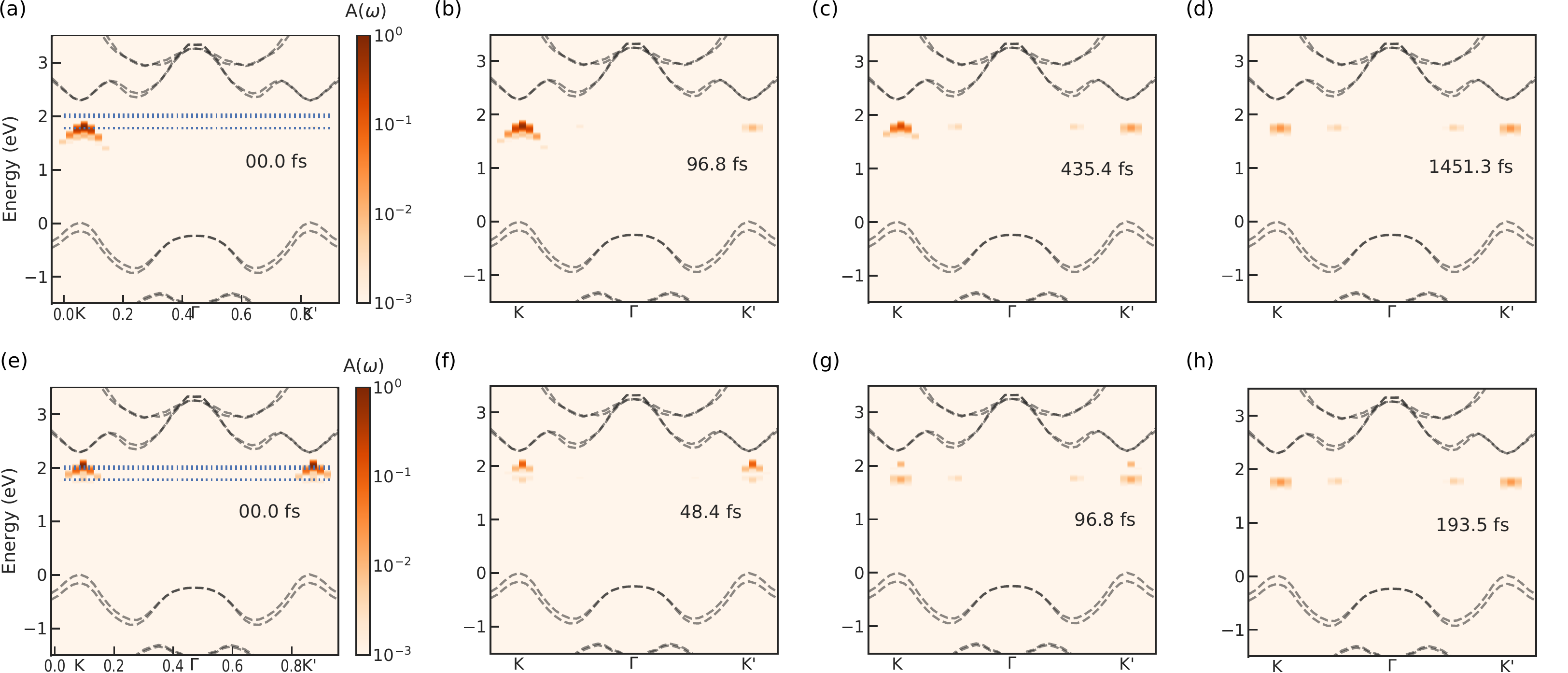}
\caption{Simulated ARPES intensity at different probe time starting with an initial A exciton state ((a),(b),(c),(d)) or a distribution of high energy excitons ((e),(f),(g),(h)) at 300 K, respectively. Dashed black lines are GW quasi-particle band dispersions and dotted horozontal blue lines indicate the energy of A exciton and two initially occupied bright exciton states in setup 3 and 4.}
\label{fig:arpes}
\end{figure*}

We simulate exciton signal intensity in TR-ARPES assuming a slowly time-varying quasi-equilibrium population of excitons using the expression~\cite{Kemper2018},
\begin{equation}
P(\omega, \vk, t) = \sum_{cv,n,\vQ}f_{n,\vQ}(t)\left|A^n_{c\vk+\vQ,v\vk}\right|^2\delta(\omega - \epsilon_{n\vQ}-\varepsilon_{v\vk}),
\label{eq:trARPES}
\end{equation}
where $A^n_{c\vk+\vQ,v\vk}$ is the $\vk$-space exciton envelope function of a state ($n$,$\mathbf{Q}$) and $\varepsilon_{v\vk}$ is the valence band electron energy with momentum $\vk$. The intensity is computed by taking exciton distribution $f_{n,\vQ}(t)$ at time $t$ from the simulations using Eq. (1). As can be seen from the argument in the delta function, for an exciton with a COM $\mathbf{Q}$ we expect that a spectral intensity disperses as replicas of valence bands of the constitutent holes, weighted by the exciton population and its envelope function square~\cite{Kemper2018,Madeo2020}, which has also been demonstrated experimentally. In Fig.~\ref{fig:arpes}, we show snapshots of intensity spectra from the 300 K simulations corresponding to Fig.~\ref{fig:popt}(e-f). The replica features can be clealy seen in Fig.~\ref{fig:arpes} (a) and (e), where we start with an A exciton, composed of electrons and holes localized in the $K$ valley and high energy excitons located at both $K$ and $K'$ valleys, respectivley. 

For a valley polarized initial state (upper panels in Fig. 4), we find that ARPES signals start to appear at $K'$ valley around 100 fs. Subsequently, the $\Lambda$ valley closest to the $K$ valley is populated around 150 fs (not shown), corresponding to the occupation of $M$ and $\Lambda$ excitons.
Experimentally, a valley transfer time of less than 500 fs to the $\Lambda$ valley is also reported in a monolayer WS$_2$~\cite{Madeo2020}.
At 435 fs, both $\Lambda$ valleys gain population but the $K$ valley exciton is still higher in population than the $K'$ valley. An equilibrium between populations in different valleys is reached around 1.5 ps. 

For the case with an initial state of higher energy excitons at both $K$ and $K'$ valleys, within 100 fs, we observe a secondary intensity maximum at lower energy in Fig.~\ref{fig:arpes} (f) and (g), which indicates the fast cooling of high energy excitons and is consistent with the quick rising in population of $\Gamma$ and $K$ excitons in Fig.~\ref{fig:popt} (f).
Similar to the case with a valley polarized initial state, we also observe a fast transfer to $\Lambda$ valley at 96.8 fs.
In Fig.~\ref{fig:arpes} (d) and (h) we show the spectrum at time when we find that the spectrum barely changes visibly. In addition to the transfer of spectral wieghts to the $\Lambda$ valley, we find the spectral dispersion flattens and is no longer concave in contrast to snapshots at earlier times. 
The flattened intensity distribution in energy can be explained by the occupation of finite momentum excitons due to the finite spreading in energy of the Boltzmann distribution.
As shown in Eq.~\ref{eq:trARPES} the total intensity is a superposition of valence band dispersion shifted in energy and momentum corresponding to exciton dispersion weighted by their populations~\cite{Kemper2018}.
At 300 K and longer time, a significant portion of higher energy and finite momentum excitons can be expected from Boltzmann distributions which leads to the flattened dispersion in Fig.~\ref{fig:arpes} (d) and (h).

\section{Thermalization}
\label{sec:thermalization}

Exciton populations are expected to reach a Bose-Einstein (BE) distribution, which should be virtually indistinguishable from the Boltzmann one at the density and temperature in our simulations, at a given temperature in the long time limit. One can indeed show that a BE distribution is a steady state solution of Eq.~\ref{eq:boltz}.
Although all simulations at the same temperature reach the same final state in the long time limit, different initial conditions lead to quite different thermalization processes and times as we will show below.
Importantly, we will show that the total thermalization, where a considerable amount of spin-anti-aligned excitons are populated, is often bottlenecked by the slow spin-flip exciton-phonon scattering process.

\begin{figure*}[t]
     \centering
     \includegraphics[width=\textwidth]{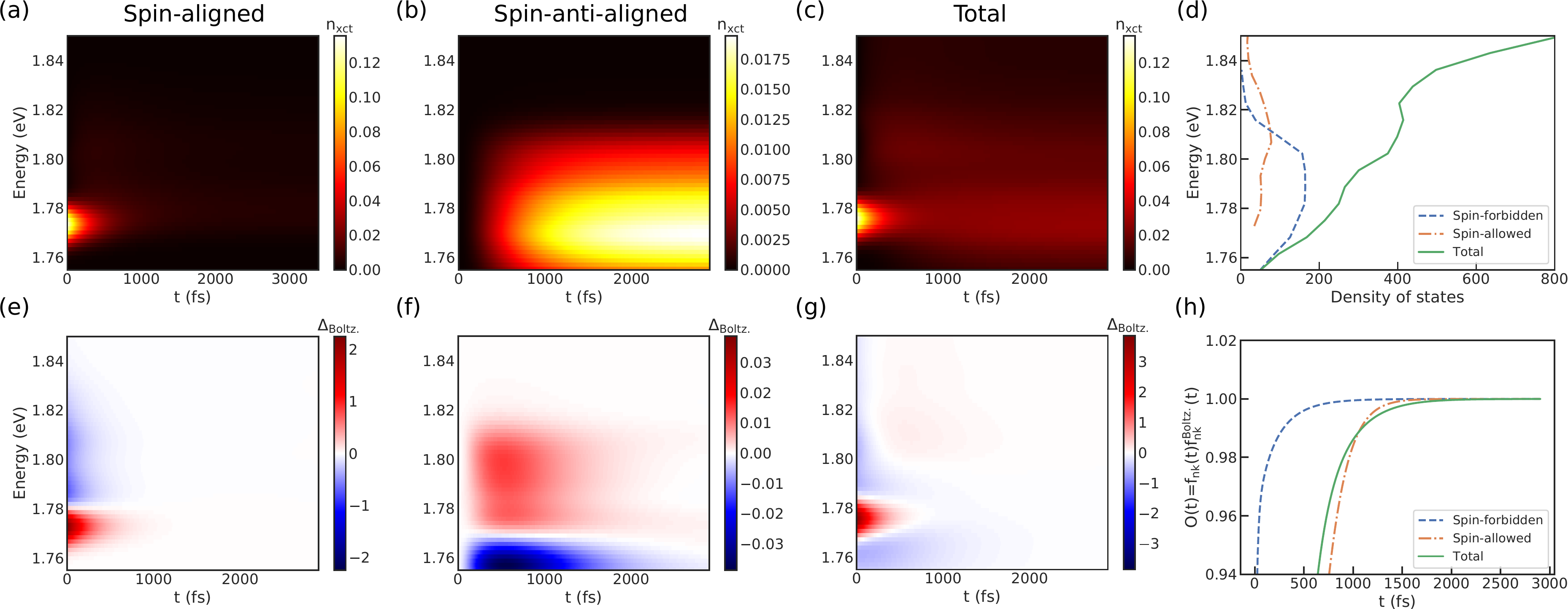}
\caption{Evolution of energy distribution of exciton populations starting with the A exciton at 300 K. Panel (a), (b), and (c) are evolutions the population of spin-aligned, spin-anti-aligned excitons at low energy valleys and of their total population, respectively. Panels (e), (f), and (g) in the bottom row are corresponding deviations from Boltzmann distributions (see main text). Note that each panel uses a different color scale. (d) Density of exciton states and (h) thermalization metrics (see text for the exact definition of $O(t)$) for spin-aligned (orange dash-dotted lines), spin-anti-aligned (blue dashed lines) at low energy valleys as defined in Fig.\ref{fig:exciton_dispersion} and total excitons (green solid lines) in the whole energy window.}
\label{fig:dist300KA}
\end{figure*}

\begin{figure*}[t]
     \centering
     \includegraphics[width=\textwidth]{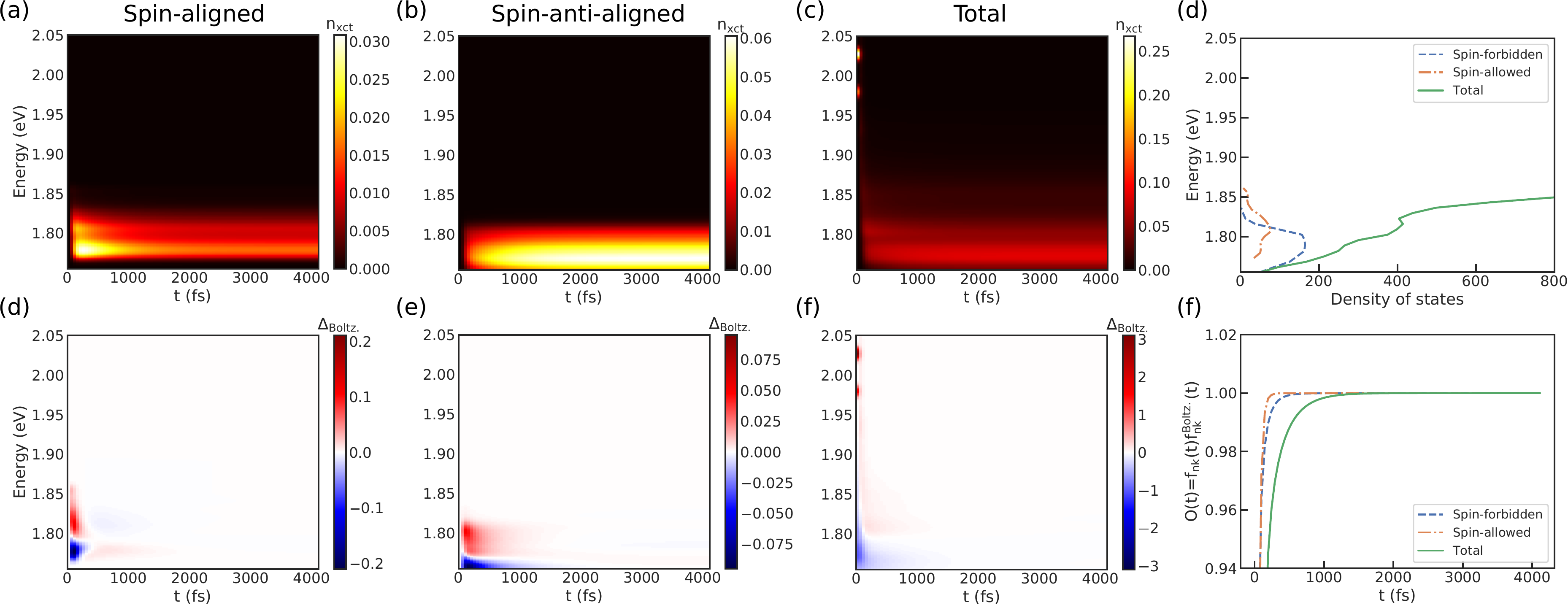}
\caption{Same as Fig.~\ref{fig:dist300KA} but the initial exciton distribution is a Gaussian distribution centered at 2.05 eV. }
\label{fig:dist300K_Ec205}
\end{figure*}

To investigate the thermalization process, we analyze energy distribution functions of excitons and their deviations from BE distributions over time in Fig.~\ref{fig:dist300KA} for setup 2, and Fig.~\ref{fig:dist300K_Ec205} for setup 4, where their population dynamics are shown in Fig.~\ref{fig:popt} (e) and (f), respectively.
Similar analysis for setup 1 and 3 are given in the Appendix.
In Fig.~\ref{fig:dist300KA} we show the evolution of energy distribution for the simulation with an initial A exciton at a fixed lattice temperature of 300 K. 
Since spin is almost a good quantum number in a large portion of the Brillouin zone, we also track the dynamics in the spin-aligned and spin-anti-aligned channels in addition to the total populations. 
We consider only low energy excitons in the spin-aligned or spin-anti-aligned channels, of which the populations are computed by summing over populations nearby the corresponding valleys as defined in Fig.~\ref{fig:exciton_dispersion}. 
The results for each channel and total population are shown in Fig.~\ref{fig:dist300KA} (a), (b), and (c) as labeled. 
We further show their normalized deviations from a Boltzmann distribution defined as $\Delta_{Boltz}(E)=(f(E)-f^{Boltz}(E))/max(f^{Boltz}(E))$ for each channel, in Fig.~\ref{fig:dist300KA} (e), (f), and (g), respectively.

Fig.~\ref{fig:dist300KA} (a) and (e) clearly show that initially only the A exciton is populated and the $\text{max}(\Delta_{Boltz})$ value of the spin-aligned channel drops below 0.1 in about 1 ps. 
Populations of low energy spin-anti-aligned excitons reach appreciable values at around 300 fs as shown in Fig.~\ref{fig:dist300KA} (b). 
During the whole process the $\Delta_{Boltz}$ in the spin-anti-aligned channel is considerably lower than that in the spin-aligned channel as shown in Fig.~\ref{fig:dist300KA} (f). 
The uneven total distribution in Fig.~\ref{fig:dist300KA} (c) is due to density of states effects as we can see from the computed exciton density of states in each channel in Fig.~\ref{fig:dist300KA} (d).

To quantify the thermalization time, we define the thermalization metric as the overlap of the exciton population and the corresponding Boltzmann distribution $O(t,T)=\sum_{n\vQ} f_{n\vQ}(t,T)*f^{Boltz}_{n\vQ}(t,T)/(\sum_{n\mathbf{Q}}f_{n\vQ}(t,T))^2$, where $t$ is the simulation time and $T$ is the temperature. The metric of each channel is defined by summing states in that channel only. For spin-anti-aligned and spin-aligned excitons, $O(t,T)$ are computed with a Boltzmann distribution given the instantaneous population in that channel at $t$. 
To be quantitative in the following discussion, we define the thermalization time as the time where the metric reaches 0.99.

From Fig.~\ref{fig:dist300KA} (h), we find that the total thermalization time is about 1 ps.   
Interestingly, we observe a quite different time scale for partial thermalization within spin-anti-aligned and spin-aligned exciton channels. 
For the spin-anti-aligned channel, thermalization is already reached within the first 100 fs.
We note that although the spin-anti-aligned exciton distribution is close to the thermalized one at 100 fs, the population in this channel is still increasing over time until the total thermalization is reached.
Such result can be understood by the fast intra-channel scattering rate and the slow inter-channel relaxation, which relies on the spin-flip scattering.
This is also consistent with the low $\Delta_{Boltz}$ value observed in Fig.~\ref{fig:dist300KA} (f).  
In contrast, it takes about 1 ps for the spin-aligned channel to reach the same metric value, which is simply due to the initially over-populated $\Gamma_3$ excitons.

Next, we investigate the thermalization process when the initial excitation is close to the quasiparticle band edge.
In Fig.~\ref{fig:dist300K_Ec205}, we perform the same analysis for the simulation in Fig.~\ref{fig:popt} (d) where the initial distribution of excitons is a Gaussian function centered at 2.05 eV with an energy width of 0.05 eV. 
At $t=0$, we can see two bright spots around 2.0 eV in Fig.~\ref{fig:dist300K_Ec205} (c), which represent the initial states. It takes less than 50 fs for excitons to populate low energy valleys, which demonstrates the fast cooling of high energy excitons. 
The fast relaxation leads to a high population of spin-aligned excitons at around 1.78 eV at 200 fs as shown in Fig.~\ref{fig:dist300K_Ec205} (a), which corresponds to the sharp rise of $K_2$ excitons observed in Fig.~\ref{fig:popt} (d). Intensity around 1.8 eV are from $\Lambda_1$ excitons. 
Popuplations of the spin-anti-aligned excitons also reach a finite value after the fast cooling of high energy excitons. Intensity located at 1.77 eV in Fig.~\ref{fig:dist300K_Ec205} (b) are mostly from $\Gamma_1$ and $\Gamma_2$ excitons.  
In both channels the $\Delta_{Boltz}$ values drop below 0.01 and are barely visible in Fig.~\ref{fig:dist300K_Ec205}  (d) and (e) within 1 ps. 
In contrast to the case in Fig.~\ref{fig:dist300KA}, overall the deviation from Boltzmann distributions in each individual channel are quite small after the first 200 fs, which can be understood from the long relaxation time of A exciton compared to those of high energy excitons.
Indeed, from the thermalization metric shown in Fig.~\ref{fig:dist300K_Ec205} (f) we find that partial thermalization is reached within 100 fs in both channels.
The total thermalization time is about 500 fs, which is shorter than the previous case.

In general, we can identify three stages of the thermalization process for monolayer MoS$_2$. First, the inital cooling of high energy excitons occurs in the shortest time scale in the whole process. After that, partial thermalization within each spin channel is established although their populations keep increasing. The whole process is complete when the total distribution reaches a Boltzmann distribution.
The separate thermalization and the long total thermarlization time are closely tied to the slow spin-flip process.

Comparing the two different setups we find that the initial excitation plays a key role in thermalization. Thermalization is faster if the initial excitation starts with high energy excitons.
We also perform the same analysis for low temperature cases (see Appendix) and find that the thermalization time is 25 ps and 20 ps for simulations in Setup 1 and 3, respectively.
The overall picture of the thermalization process at 100 K is similar to that at 300 K but the time scale is one order of magnitude longer.

\section{Conclusion}
\label{sec:conclusion}

In conclusion, our first-principles simulation of exciton thermalization dynamics in monolayer MoS$_2$ with exciton-phonon coupling reveals interesting microscopic details of the thermalization process. 
We focus on the dynamics following either a resonant excitation of the A exciton or a near-band-edge exciation of a distribution of excited excitons at 100 K and 300 K. 
We find that the thermalization time strongly depends on the temperature (an order of magnitude difference can already be seen at these two temperatures), but qualitatively, the thermalization pathways remain similar for similar initial exciton populations, regardless of the temperature, at least for the cases examined here.
Qualitative changes to the thermalization occur for different initial excitations.
For resonant excitations, the slow A exciton relaxation time leads to an interesting population trapping behavior, which is enhanced at low temperatures.
Moreover, through a detailed comparison with the Boltzmann distribution, we identify different stages of thermalization. 
We find that spin being an almost good quantum number leads to a clear separation of specific thermalization time for individual spin channels and total thermalization time.
The momentum and energy-resolved features we predict may be observed via TR-ARPES experiments. 
Our simulated TR-ARPES results indicate a fast valley transfer time of less than 100 fs, which agrees reasonably with previous experiments.
Our results lay a detailed foundation for understanding the relationship between the initial excitation, thermalization pathways, and experimental signatures of exciton dynamics.

\section*{Acknowledgments}
\label{sec:acknowledgments}

This work was primarily supported by the Center for Computational Study of Excited State Phenomena in Energy Materials (C2SEPEM), which is funded by the U.S. Department of Energy, Office of Science, Basic Energy Sciences, Materials Sciences and Engineering Division under Contract No. DE-AC02-05CH11231, as part of the Computational Materials Sciences Program.
We acknowledge the use of computational resources at the National Energy Research Scientific Computing Center (NERSC), a DOE Office of Science User Facility supported by the Office of Science of the U.S. Department of Energy under Contract No. DE-AC02-05CH11231. The authors acknowledge the Texas Advanced Computing Center (TACC) at The University of Texas at Austin for providing HPC resources that have contributed to the research results reported within this paper. YHC thanks the National Center for High-Performance Computing (NCHC) for providing computational and storage resources.

\section*{Appendix}
\label{sec:appendix}

In this appendix, we include two simulation results similar to those in the main text but the temperature is set at 100 K.
We find that qualitatively the thermalization dynamics is similar to that at 300 K but the thermalization time scale is much longer.
The separation of thermalization in each spin channel and the long thermalization time due to spin-flip scattering can be clearly identified.

\begin{figure*}[t]
     \centering
    \includegraphics[width=\textwidth]{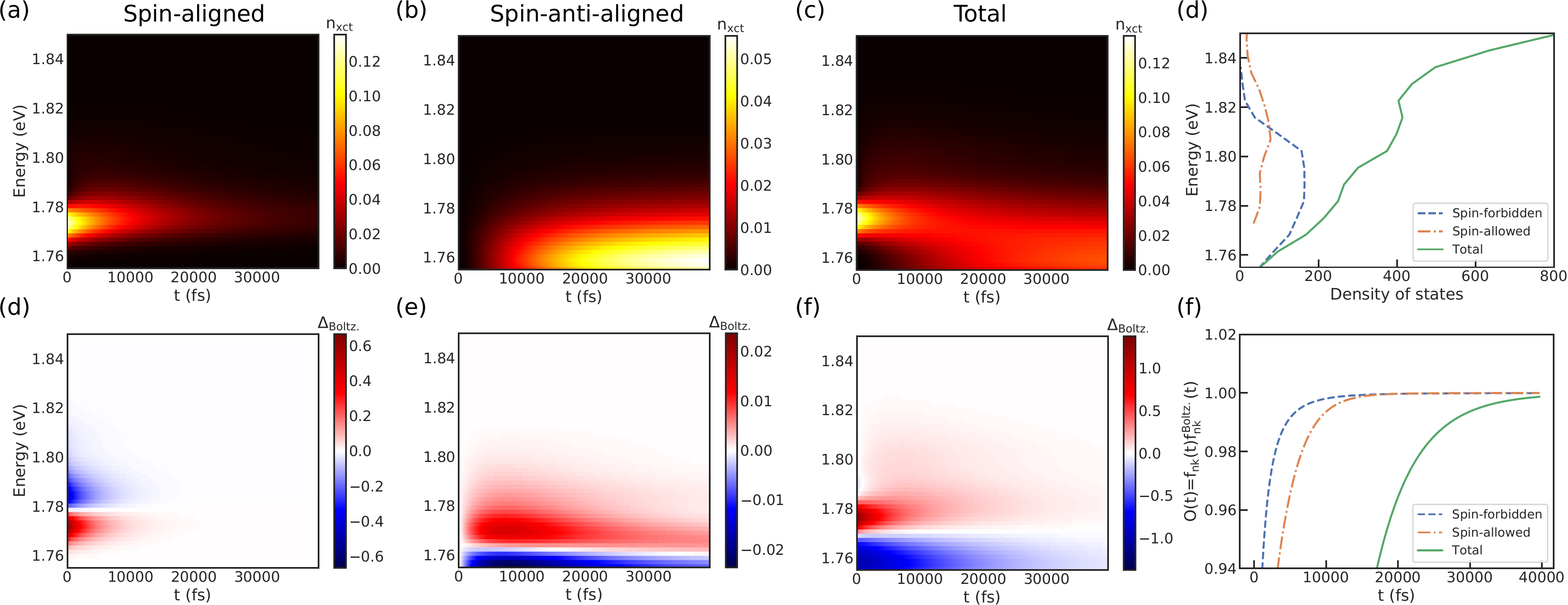}
\caption{Evolution of energy distribution of exciton populations starting with A exciton at 100
K. Panel (a), (b), and (c) are evolutions of spin-aligned, spin-anti-aligned exciton at low energy
valleys and of total populations, respectively. Panels (e), (f), and (g) in the bottom row are
corresponding deviations from Boltzmann distributions (see main text). (d) Density of states and
(h) thermalization metrics for spin-aligned (orange dash-dotted lines), spin-anti-aligned (blue dashed
lines) and total excitons (green solid lines)}
\label{fig:dist100KA}
\end{figure*}

\begin{figure*}[t]
      \centering
      \includegraphics[width=\textwidth]{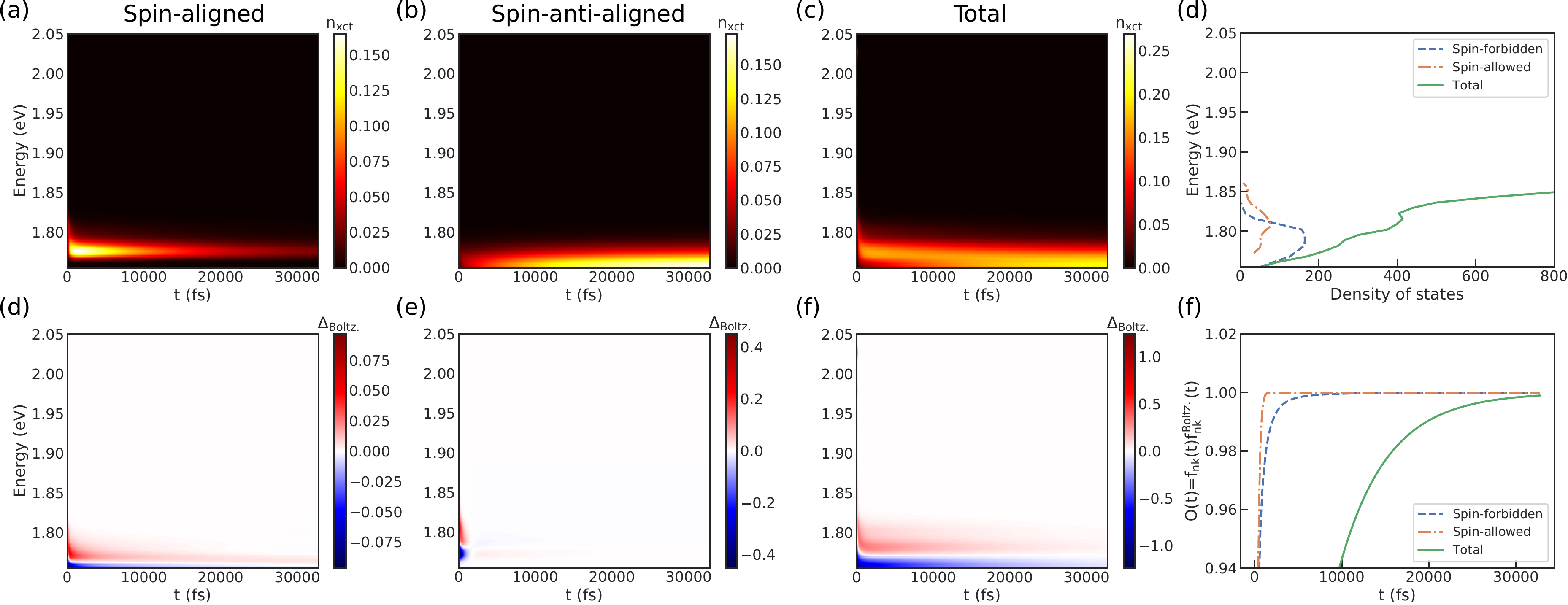}
 \caption{Evolution of energy distribution of exciton populations with an initial Gaussian distribution of excitons centered at 2.05 eV at 100 K. Panel (a), (b), and (c) are evolutions of spin-aligned, spin-anti-aligned exciton at low energy
valleys and of total populations, respectively. Panels (e), (f), and (g) in the bottom row are
corresponding deviations from Boltzmann distributions (see main text). (d) Density of states and
(h) thermalization metrics for spin-aligned (orange dash-dotted lines), spin-anti-aligned (blue dashed
lines) and total excitons (green solid lines)}
 \label{fig:dist100K_Ec205}
\end{figure*}

\section*{References}
\bibliography{ref}

\end{document}